\documentclass[prb,aps,showpacs,floatfix,floats,nobalancelastpage,twocolumn]{revtex4}
\usepackage{graphicx}
\usepackage{amsmath}
\usepackage{color}
\usepackage{dcolumn}
\usepackage{bm}
\usepackage{epstopdf}

\def\etal{~\textit{et~al.}}

\def\up{\uparrow}
\def\dn{\downarrow}
\def\Hc{{\rm H.c.}}

\def\ET{{$\kappa$-(ET)$_2$Cu$_2$(CN)$_3$}}
\def\dmit{{EtMe$_3$Sb[Pd(dmit)$_2$]$_2$}}
\def\hyperkag{{Na$_4$Ir$_3$O$_8$}}

\begin{document}

\title{Two-band electronic metal and neighboring spin liquid
(spin Bose-metal) on a zigzag strip with longer-ranged repulsion}
\author{Hsin-Hua Lai}
\affiliation{Department of Physics, California Institute of Technology, Pasadena, California 91125}
\author{Olexei I. Motrunich}
\affiliation{Department of Physics, California Institute of Technology, Pasadena, California 91125}
\date{\today}

\begin{abstract}
We consider an electronic model for realizing the
Spin Bose-metal (SBM) phase on a 2-leg triangular strip --
a spin liquid phase found by D.~N.~Sheng\etal
[Phys. Rev. B {\bf 79}, 205112 (2009)]
in a spin-1/2 model with ring exchanges.
The SBM can be viewed as a ``C1S2'' Mott insulator of electrons
where the overall charge transporting mode is gapped out.
We start from a two-band ``C2S2'' metal and consider extended repulsion
motivated by recent ab initio derivation of electronic model for
$\kappa$-ET spin liquid material
[K.~Nakamura\etal, J. Phys. Soc. Jpn. {\bf 78}, 083710(2009)].
Using weak coupling renormalization group analysis, we find that
the extended interactions allow much wider C2S2 metallic phase
than in the Hubbard model with on-site repulsion only.
An eight-fermion Umklapp term plays a crucial role in producing
a Mott insulator but can not be treated in weak coupling.
We use Bosonization to extend the analysis to intermediate coupling
and study phases obtained out of the C2S2 metal upon increasing overall
repulsion strength, finding that the SBM phase is a natural outcome
for extended interactions.
\end{abstract}
\maketitle

\section{Introduction}

There has been much recent interest in gapless spin liquids
stimulated by the appearance of several experimental candidates,
including two-dimensional (2D) triangular lattice based organic
compounds\cite{Shimizu03, Kurosaki05, SYamashita08, MYamashita09, Itou08}
\ET\ and \dmit\ and 3D hyper-kagome material\cite{Okamoto07} \hyperkag.
One line of theoretical ideas considers states with a Fermi surface of
fermionic spinons.\cite{ringxch, SSLee, Zhou_hypkag, Lawler_hypkag}
For the 2D spin liquids, such a state arises as a good variational
wavefunction\cite{ringxch} for an appropriate spin model with ring
exchanges; it is also an appealing candidate for the Hubbard model
near the Mott transition.\cite{SSLee, Senthil_Mott, Podolsky09}
Theoretical description of such states leads to a U(1) gauge theory
(see Ref.~\onlinecite{LeeNagaosaWen} for a review).

Variational studies are not sufficient to prove that a given state
is realized and the gauge theory is not fully reliable in 2D.
Driven by the need for a controlled theoretical access to such phases,
Ref.~\onlinecite{Sheng09} considered the Heisenberg plus ring exchanges
model on a two-leg triangular strip (so-called zigzag chain).
Using numerical Density Matrix Renormalization Group (DMRG),
Variational Monte Carlo (VMC), and analytical Bosonization treatment,
Ref.~\onlinecite{Sheng09} found a ladder descendant of the 2D spin liquid
in a broad range of parameters and dubbed this phase ``Spin Bose-Metal''
(SBM).  The name refers to metal-like itinerancy present in the spin
degrees of freedom, while there is no electric transport to speak of
in the spin-only model, which is bosonic model microscopically.

A low energy field theory\cite{Sheng09} for the zigzag SBM phase can be
obtained by employing bosonization to analyze the spinon-gauge theory
(the slave particle approach also underlies the VMC trial states).
An alternative derivation of the SBM theory is to consider an
interacting model of {\it electrons} hopping on the zigzag chain and to
drive a transition from a two-band metal to a particular
Mott insulator.  Specifically, let us start in the metallic phase with
two gapless charge modes and two gapless spin modes -- so-called
``C2S2'' metal.  We can imagine gapping out just the overall charge mode
to obtain a ``C1S2'' Mott insulator with one gapless ``charge'' mode and
two gapless spin modes, where the former represents local current loop
fluctuations and does not transport charge along the chain.
This is precisely the SBM phase.  If one thinks of a spin-only
description of this Mott insulator, the gapless ``charge'' mode can be
interpreted as spin singlet chirality mode.
Ref.~\onlinecite{Sheng09} also identified a valid Umklapp term that
can drive the electron system to the C1S2 phase.

In this paper, we focus on realizing such scenario for the SBM in
explicit and realistic electronic models.
Hubbard model on the zigzag chain ($t_1 - t_2 - U$ chain) has received
much attention.\cite{Fabrizio96, Louis01, Gros05, Daul, Fabrizio97, Japaridze}
For free electrons, the two-band metal appears for $t_2/t_1 > 0.5$.
However, in the case of Hubbard interaction, weak coupling approach
\cite{Louis01, Gros05} finds that this phase is stable only over a
narrow range $t_2/t_1 \in [0.5, 0.57]$, while a spin gap opens up
for larger $t_2/t_1$.
The Umklapp that can drive a transition to a Mott insulator requires
eight fermions and is strongly irrelevant at weak coupling.
Prior work\cite{Balents96, Fabrizio96, Louis01} focused on the
spin-gapped metal and eventual spin-gapped insulator for strong
interaction, while the C1S2 spin liquid phase was not anticipated.

There have also been numerical DMRG studies of the Hubbard model.\cite{Daul, Fabrizio97, Japaridze, Gros05}
The focus has been on the prominent spin-gapped phases and, in particular,
on the insulator that is continuously connected to the dimerized phase
in the $J_1 - J_2$ Heisenberg model, which is appropriate in the
strong interaction limit $U \gg t_1, t_2$.
The C2S2 metallic phase and possibility of nearby spin liquid on the
Mott insulator side in the Hubbard model have not been explored.
We hope our work will motivate more studies of this interesting
possibility in the Hubbard model with intermediate $U$ close to the
C2S2 metal.

Since the C2S2 metallic phase is quite narrow in the Hubbard model,
we would like to first widen the C2S2 region.  To this end, we explore
an electronic model with extended repulsive interactions.\cite{Kane97}
Such interactions tend to suppress instabilities in the electronic
system, similar to how long-ranged Coulomb repulsion suppresses
pairing in metals.  They are also more realistic than the
on-site Hubbard, particularly for materials undergoing a metal-insulator
transition where there is no conduction band screening on the
insulator side.  Thus, recent ab initio model construction for the \ET\
material found significant extended interactions in the corresponding
electronic model on the half-filled triangular lattice.\cite{Nakamura09, Valenti09}

Applying weak coupling renormalization group (RG) approach to the
zigzag ladder system,\cite{Balents96, Fabrizio96, Lin97, Louis01}
we indeed find that extended interactions open a much wider window
of the C2S2 metal phase.
Building on this, we then use bosonization approach to explore
a transition to a Mott insulator upon increasing the overall repulsion
strength.  We find that such longer-ranged interactions can
drive the system into the C1S2 spin liquid Mott insulator
rather than a spin-gapped insulator.
This bodes well for finding spin liquid phases in more realistic
electronic models for materials near the metal-insulator transition.

The paper is organized as follows.
In Sec.~\ref{sec:Weak coupling}, we set up the weak coupling RG
\cite{Balents96, Lin97, Louis01} and open a much wider window of the
metallic C2S2 phase by introducing realistically motivated
longer-ranged repulsion.
In Sec.~\ref{sec:Intermediate coupling}, we use bosonization to
extend the analysis to intermediate coupling.  We gradually increase the
overall repulsion strength and determine thresholds for a Mott
transition driven by the eight-fermion Umklapp term and also for spin gap
instabilities, thus mapping out phases neighboring the C2S2 metal.
In Sec.~\ref{conclusion}, we summarize our results and conclude
with some discussion.

\section{Weak coupling analysis of $t_1 - t_2$ model with extended
repulsion: Stabilizing C2S2 metal}
\label{sec:Weak coupling}

\subsection{Setup for two-band electron system}

We consider half-filled electronic $t_1 - t_2$ model with extended
interaction described by the Hamiltonian $H = H_0 + H_V$, with
\begin{eqnarray}
\nonumber
&& H_0 = -\sum_{x,\alpha} \big[ t_1 c^\dagger_\alpha(x) c_\alpha(x+1)
+ t_2 c^\dagger_\alpha(x) c_\alpha(x+2) \\
&& \hspace{6cm} + \Hc \big] ~, \label{freehamiltonian} \\
&& H_V = \frac{1}{2} \sum_{x,x'} V(x-x') n(x) n(x') ~.
\label{long-ranged repulsion}
\end{eqnarray}
Here $c (c^\dagger)$ is fermion annihilation (creation) operator,
$x$ is a site label on the one-dimensional (1D) chain, and
$\alpha = \uparrow, \downarrow$ is a spin index;
$n(x) \equiv \sum_\alpha c^\dagger_\alpha(x) c_\alpha(x)$ is
electron number on the site.

\begin{figure}
   \centering
   \includegraphics[width=\columnwidth]{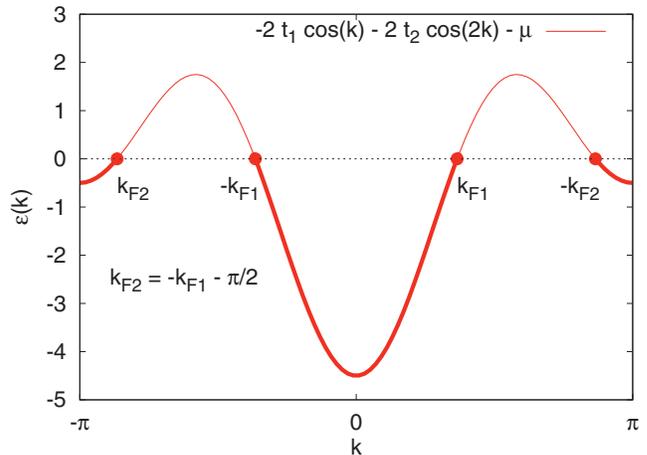}
   \caption{Electron band for $t_2 > 0.5 t_1$ has two occupied
Fermi sea segments.  This is free fermion C2S2 metal.}
   \label{energy dispersion}
\end{figure}

In weak coupling, the kinetic energy Eq.~(\ref{freehamiltonian})
gives free particle dispersion
\begin{eqnarray}
\epsilon(k) = -2 t_1 \cos(k) - 2 t_2 \cos(2k) ~.
\end{eqnarray}
For $t_2/t_1 > 0.5$, there are two sets of Fermi points at wavevectors
$\pm k_{F1}$ and $\pm k_{F2}$ as shown in Fig.~\ref{energy dispersion}.
We adopt the same conventions as in Ref.~\onlinecite{Sheng09}.
Fermions near $k_{F1}$ and $k_{F2}$ are moving to the right, and the
corresponding group velocities are $v_1, v_2 > 0$.
Electrons are at half-filling, which implies
$k_{F1} + k_{F2} = -\pi/2 \mod 2\pi$ for the choices as in
Fig.~\ref{energy dispersion}.

The electron operators are expanded in terms of continuum fields,
\begin{eqnarray}
c_\alpha(x) = \sum_{P,a} e^{i P k_{Fa} x} c_{Pa\alpha} ~,
\end{eqnarray}
with $P = R/L = +/-$ denoting the right and left movers and
$a = 1, 2$ denoting the two Fermi seas.

Four-fermion interactions can be conveniently expressed in terms of
chiral currents,\cite{Balents96, Lin97, Sheng09}
\begin{eqnarray}
J_{Pab} &=& \sum_\alpha c_{Pa\alpha}^\dagger c_{Pb\alpha} ~,
\label{charge current} \\
\vec{J}_{Pab} &=& \sum_{\alpha, \beta}
c_{Pa\alpha}^\dagger \frac{\vec{\sigma}_{\alpha \beta}}{2} c_{Pb\beta}
~.
\label{spin current}
\end{eqnarray}
Most general four-fermion interactions can be written as,
\begin{eqnarray}
\mathcal{H}^\rho_{RL} &=& \sum_{a,b}
\big( w^\rho_{ab} J_{Rab} J_{Lab} + \lambda^\rho_{ab} J_{Raa} J_{Lbb}
\big) ~, \label{rlrhohamiltonian} \\
\mathcal{H}^\sigma_{RL} &=& -\sum_{a,b}
\big( w^\sigma_{ab} \vec{J}_{Rab} \cdot \vec{J}_{Lab}
     + \lambda^\sigma_{ab} \vec{J}_{Raa} \cdot \vec{J}_{Lbb}
\big) ~, \label{rlsigmahamiltonian} \\
\nonumber \mathcal{H}^{\rho}_{\rm chiral} &=& \frac{1}{2}\sum_{a}C^{\rho}_{aa}\big( J_{Raa}J_{Raa}+J_{Laa}J_{Laa}\big) \\
&+& C^{\rho}_{12} \big( J_{R11}J_{R22}+J_{L11}J_{L22}\big)~, \label{chiralrhohamiltonian}\\
\nonumber \mathcal{H}^\sigma_{\rm chiral} &=& -\frac{1}{2} \sum_a C^\sigma_{aa}
\big( \vec{J}_{Raa} \cdot \vec{J}_{Raa}
      + \vec{J}_{Laa} \cdot \vec{J}_{Laa} \big) \\
&-& C^\sigma_{12} \big( \vec{J}_{R11} \cdot \vec{J}_{R22}
                        + \vec{J}_{L11} \cdot \vec{J}_{L22} \big) ~.
\label{chiralsigmahamiltonian}
\end{eqnarray}
Here $\mathcal{H}_{RL}$ are terms that connect right and left movers,
while $\mathcal{H}_{\rm chiral}$ are chiral terms with all fermions
moving in the same direction.

Consider the couplings in $\mathcal{H}_{RL}$.  We have
$w_{11} = w_{22} = 0$ (convention), $w_{12} = w_{21}$ (from Hermiticity),
and $\lambda_{12} = \lambda_{21}$ (from $R \leftrightarrow L$ symmetry).
Thus there are 8 independent couplings:
$w^{\rho/\sigma}_{12}, \lambda^{\rho/\sigma}_{11},
\lambda^{\rho/\sigma}_{22}$, and $\lambda^{\rho/\sigma}_{12}$.
Note that there are no four-fermion Umklapp terms in our two-band system.

In the specific lattice model, we expand the interactions
Eq.~(\ref{long-ranged repulsion}) in terms of the continuum fields
and find ``bare'' values of the couplings:
\begin{eqnarray}
\lambda^\rho_{11} &=& V_{Q=0} - \frac{V_{2k_{F1}}}{2} ~,\label{rho11}\\
\lambda^\rho_{22} &=& V_{Q=0} - \frac{V_{2k_{F2}}}{2} ~,\label{rho22}\\
\lambda^\rho_{12} &=& V_{Q=0} - \frac{V_{\pi/2}}{2} ~, \label{rho12}\\
\lambda^\sigma_{11} &=& 2 V_{2k_{F1}} ~, \label{sigma11}\\
\lambda^\sigma_{22} &=& 2 V_{2k_{F2}} ~, \label{sigma22}\\
\lambda^\sigma_{12} &=& 2 V_{\pi/2} ~, \label{sigma12}\\
w^\rho_{12} &=& V_{k_{F1}-k_{F2}} - \frac{V_{\pi/2}}{2} ~,
\label{wrho12}\\
w^\sigma_{12} &=& 2 V_{\pi/2} ~, \label{wsigma12}\\
C^\rho_{11} &=& C^\rho_{22} \,=\, V_{Q=0} - \frac{U}{2} ~,
\label{crho11} \\
C^\rho_{12} &=& V_{Q=0} - \frac{V_{k_{F1}-k_{F2}}}{2} ~, \label{crho12}\\
C^\sigma_{11} &=& C^\sigma_{22} \,=\, 2 U ~, \label{csigma11} \\
C^\sigma_{12} &=& 2 V_{k_{F1} - k_{F2}} ~. \label{csigma12}
\end{eqnarray}
Here
$V_Q \equiv \sum_{x' = -\infty}^\infty V(x-x') e^{i Q (x-x')} = V_{-Q}$,
since $V(x-x') = V(x'-x)$.
We have also used explicitly $k_{F1} + k_{F2} = -\pi/2$.

The terms ${\mathcal H}_{\rm chiral}$ renormalize ``velocities'' of
various modes.  In the weak coupling RG analysis, they only generate
higher order contributions and are therefore not important.
The RG equations below contain only couplings from ${\mathcal H}_{RL}$.
On the other hand, the chiral interactions are important in the
intermediate coupling analysis to be done in Sec.~\ref{sec:Intermediate coupling},
which is why we have listed their values as well.
The on-site coupling $U \equiv V(x-x' = 0)$ appears explicitly in
$C^{\rho/\sigma}_{11}$ and $C^{\rho/\sigma}_{22}$ because of our
more careful treatment of the on-site interaction, which we first
write as $U n_\up(x) n_\dn(x)$ and then insert the continuum fields
(and bosonize in Sec.~\ref{sec:Intermediate coupling}).

\subsection{Weak coupling Renormalization Group}
\label{subsec:RGeqs}

The RG equations in the two-band system are:\cite{Balents96, Lin97, Louis01}
\begin{eqnarray}
&&\hspace{-0.8cm}\dot{\lambda}^{\rho}_{11} = -\frac{1}{2\pi v_2} \left[ \left(w^{\rho}_{12}\right)^2+\frac{3}{16}\left(w^{\sigma}_{12}\right)^2\right],\label{rglambdarho11}\\
&&\hspace{-0.8cm}\dot{\lambda}^{\rho}_{22} = -\frac{1}{2\pi v_1} \left[ \left(w^{\rho}_{12}\right)^2+\frac{3}{16}\left(w^{\sigma}_{12}\right)^2\right],\label{rglambdarho22}\\
&&\hspace{-0.8cm}\dot{\lambda}^{\rho}_{12} = \frac{1}{\pi (v_1+v_2)} \left[ \left(w^{\rho}_{12}\right)^2+\frac{3}{16}\left(w^{\sigma}_{12}\right)^2\right],\label{rglambdarho12}\\
&&\hspace{-0.8cm}\dot{\lambda}^{\sigma}_{11} = -\frac{1}{2\pi v_1}\left(\lambda^{\sigma}_{11} \right)^2 -\frac{1}{4\pi v_2}\left[ \left( w^{\sigma}_{12}\right)^2+4w^{\rho}_{12}w^{\sigma}_{12}\right],\label{rglambdasigma11}\\
&&\hspace{-0.8cm}\dot{\lambda}^{\sigma}_{22} = -\frac{1}{2\pi v_2} \left(\lambda^{\sigma}_{22}\right)^2 -\frac{1}{4\pi v_1}\left[ \left(w^{\sigma}_{12}\right)^2+4w^{\rho}_{12}w^{\sigma}_{12}\right],\label{rglambdasigma22}\\
&&\hspace{-0.8cm} \dot{\lambda}^{\sigma}_{12} = -\frac{1}{\pi(v_1+v_2)}\left\{\left(\lambda^{\sigma}_{12}\right)^2+\frac{\left( w^{\sigma}_{12}\right)^2-4w^{\rho}_{12}w^{\sigma}_{12}}{2}\right\},\label{rglambdasigma12}\\
&&\hspace{-0.8cm}\dot{w}^{\rho}_{12} = -\Lambda^{\rho}w^{\rho}_{12}-\frac{3}{16} \Lambda^{\sigma}w^{\sigma}_{12} ~,\label{rgwrho12}\\
&&\hspace{-0.8cm}\dot{w}^{\sigma}_{12} = -\Lambda^{\sigma}w^{\rho}_{12}-\left(\Lambda^{\rho}+\frac{\Lambda^{\sigma}}{2}+\frac{2 \lambda^{\sigma}_{12}}{\pi(v_1 + v_2)}\right) w^{\sigma}_{12} ~.\label{rgwsigma12}
\end{eqnarray}
Here $\dot{O}\equiv \partial{O}/\partial{\ell}$, where $\ell$ is
logarithm of the length scale.  We have also defined
\begin{equation}
\label{c2s2instability}
\Lambda^{\rho/\sigma} = \frac{\lambda^{\rho/\sigma}_{11}}{2\pi v_1}
+ \frac{\lambda^{\rho/\sigma}_{22}}{2\pi v_2}
- \frac{2\lambda^{\rho/\sigma}_{12}}{\pi (v_1 + v_2)} ~.
\end{equation}
Details of our system enter through the band velocities $v_1, v_2$,
and the initial conditions Eqs.~(\ref{rho11})-(\ref{wsigma12}).

\subsection{Fixed point for stable C2S2 phase}
\label{stable fixed point}

We are primarily interested in the stability of the two-band
metallic phase with two gapless charge and two gapless spin modes
-- ``C2S2'' in the notation of Ref.~\onlinecite{Balents96}.
In the RG, this phase is characterized as having no divergent couplings.
Before proceeding with detailed numerical studies of the flow
Eqs.~(\ref{rglambdarho11})-(\ref{rgwsigma12}),
we can describe such stable C2S2 fixed point qualitatively:
The charge sector couplings reach some fixed values,
$\lambda^{\rho*}_{11}, \lambda^{\rho*}_{22}, \lambda^{\rho*}_{12}$,
and are strictly marginal; they also need to satisfy
$\Lambda^{\rho*} > 0$ (see below).
The spin sector couplings approach zero from positive values,
$\lambda^{\sigma*}_{11} = \lambda^{\sigma*}_{22} = \lambda^{\sigma*}_{12}
= 0^+$, and are marginally irrelevant.
Finally, the ``charge-spin'' couplings $w_{12}$ go to zero,
$w^{\rho*}_{12} = w^{\sigma*}_{12} = 0$,
and are irrelevant, which is insured by the condition
$\Lambda^{\rho*} > 0$.
Indeed, consider small deviations of comparable magnitudes for
all couplings and allowing only positive $\lambda^\sigma_{ab}$.
Since we have finite $\Lambda^{\rho*} > 0$, first the
$w_{12}^{\rho/\sigma}$ will renormalize quickly to zero,
without affecting significantly the other couplings.
Then the $\lambda^\sigma_{ab}$ will renormalize to zero via slow
marginal flows.

\subsection{Numerical studies of the flows}
\label{numerical studies of flows}

We can solve the RG equations numerically for given initial conditions
and check whether the couplings flow into the domain of attraction
of the C2S2 fixed point or not.  We use Mathematica to solve the flows
up to long ``time'' $\ell$ when the ultimate trends become apparent.

If the couplings always remain of the same order as their initial values
or approach zero, we say the couplings are marginal or irrelevant and
identify this as the C2S2 phase.  The eventual trends here were discussed
in Sec.~\ref{stable fixed point}.

On the other hand, if the magnitudes of some couplings grow
significantly compared to the initial values, we say that the couplings
are relevant and the C2S2 phase is destroyed.  Thus, if either
$\lambda^\sigma_{11}$ or $\lambda^\sigma_{22}$ coupling becomes negative
while $w^{\sigma}_{12}$ and $w^{\rho}_{12}$ remain of the same sign,
this $\lambda^\sigma$ then runs away to large negative values and also
induces the other couplings to diverge.  Bosonizing the four-fermion
interactions\cite{Balents96, Lin97, Sheng09}
(cf.\ Sec.~\ref{sec:Intermediate coupling}),
we can see that two spin modes and one charge mode become gapped and
we obtain so-called ``C1S0'' phase.  The overall charge propagation mode
remains gapless and the system is conducting.
Note that we do not distinguish which coupling diverges faster in the
formal flow Eqs.~(\ref{rglambdarho11})-(\ref{rgwsigma12}).
As discussed in Ref.~\onlinecite{Balents96},
in the $U \to 0+$ limit one can separate a so-called ``C2S1'' case where
one of the spin couplings diverges qualitatively faster (but all
couplings still diverge at the same $\ell$).  We do not make such subtle
distinction and call any runaway flow situation as C1S0 -- all we
want to know is that the two-band metal C2S2 became unstable.

\begin{figure}[t]
  \centering
  \includegraphics[width=\columnwidth]{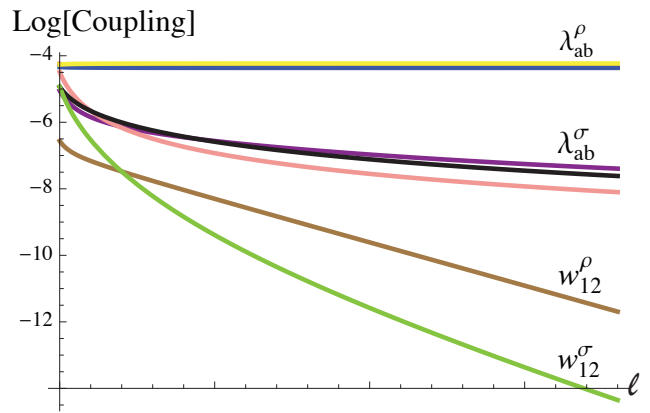}
  \caption{
(Color online)
Example of RG flows in the C2S2 phase.  The model potential is
Eq.~(\ref{parameterization}) with $\kappa=1/2$, $\gamma=2/5$;
the band parameter is $t_2/t_1 = 0.9$.
We choose {\it logarithm} of the couplings to be the $y$-axis
and RG ``time'' $\ell$ to be the $x$-axis.
We see that $w^{\rho/\sigma}_{12}$ flow toward 0 rapidly
(irrelevant couplings);
$\lambda^{\rho}_{ab}$ saturate very fast (strictly marginal couplings);
while $\lambda^{\sigma}_{ab}$ flow to 0 slowly (marginally irrelevant).
More generally, if we fix these $\kappa$ and $\gamma$ values,
for $t_2/t_1 < 0.99$ the flows are similar to those shown here
and the phase is C2S2.
}
  \label{c2s2RGflow}
\end{figure}

The RG flows are qualitatively similar for different points in the
same phase, so we only show one representative picture for each case.
Fig.~\ref{c2s2RGflow} shows the flows in the C2S2 phase.
The scale parameter $\ell$ is the $x$-axis, while logarithm of the
couplings is the $y$-axis.  In this way, we clearly see that the couplings
separate into three groups, which is well explained by the C2S2 fixed
point in Sec.~\ref{stable fixed point}:
the $w_{12}^{\rho/\sigma}$ flow to 0 exponentially rapidly,
the $\lambda^{\sigma}_{ab}$ flow to 0 marginally slowly, while
the $\lambda^{\rho}_{ab}$ saturate.

\begin{figure}[t]
  \centering
  \includegraphics[width=\columnwidth]{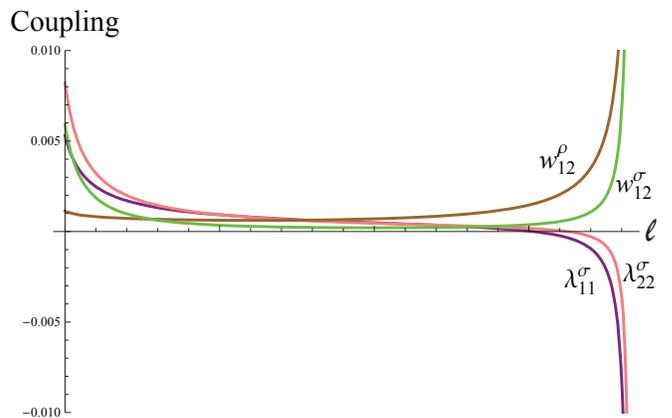}
  \caption{
(Color online)
Example of RG flows of selected couplings in the C1S0 phase.  The model
is the same as in Fig.~\ref{c2s2RGflow}, but with $t_2/t_1 = 1.05$.
We see that the selected couplings diverge after some time.  For example,
once the $\lambda^{\sigma}_{11}$ and $\lambda^{\sigma}_{22}$ become
negative while $w^{\rho}_{12}$ and $w^{\sigma}_{12}$ remain positive,
the RG equations~(\ref{rglambdarho11})-(\ref{rgwsigma12}) drive
the $\lambda^{\sigma}_{11}$ and $\lambda^{\sigma}_{22}$ to -$\infty$
and in turn $w^{\rho}_{12}$ and $w^{\sigma}_{12}$ to +$\infty$,
and then all couplings diverge.
More generally, if we fix $\gamma=2/5$, for $t_2/t_1 > 0.99$ the flows
are similar to those shown here and we call this C1S0 phase.
Varying $\gamma$, we obtain the phase diagram Fig.~\ref{extendedc2s2}.
}
  \label{c1s0RGflow}
\end{figure}

Fig.~\ref{c1s0RGflow} illustrates the flows in the C1S0 phase.
Here we use real values of the coupling as the $y$-axis and only show
selected couplings, $\lambda^{\sigma}_{11}$, $\lambda^{\sigma}_{22}$,
$w^{\rho}_{12}$, and $w^{\sigma}_{12}$.
We clearly see that these couplings diverge (and so do the other
couplings not shown in the figure).

\subsection{Examples of phase diagrams with C2S2 metal stabilized by
extended interactions}
\label{subsec:RG:numphased}

For illustration in our paper, we consider the following interaction
potential,
\begin{eqnarray}
\label{parameterization}
V(x-x') = \left\{
\begin{array}{cc}
U &, \hspace{0.5cm} |x-x'|=0 \\
\kappa U e^{-\gamma |x-x'|} &, \hspace{0.5cm} |x-x'| \geq 1
\end{array}
\right\}
\end{eqnarray}
Here $U$ is the overall energy scale and also the on-site repulsion.
The relative magnitude of the extended repulsion is set by some
factor $\kappa < 1$.  Beyond one lattice spacing, the potential
decreases exponentially with decay rate $\gamma$.
For $\gamma \to \infty$ we obtain the Hubbard model with on-site
interaction only, while for small $\gamma$ the interaction extends
over many lattice sites.

We also consider the above potential but truncated at the 4-th neighbor.
This tests robustness of our conclusions to modifications where the
interactions have finite but still somewhat extended range,
as may be preferable in numerical studies of such electronic models.

\subsubsection{Weak coupling phase diagram for potential
Eq.~(\ref{parameterization})}
\label{subsubsec:extendedc2s2}

\begin{figure}[t]
   \centering
   \includegraphics[width=\columnwidth]{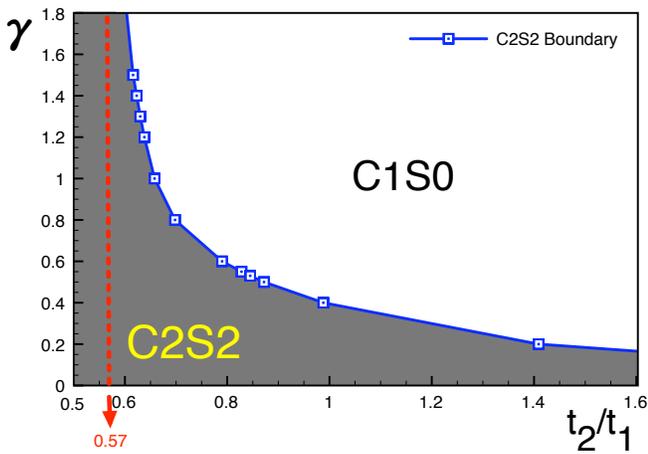}
   \caption{
Stabilization of the C2S2 metal by extended interactions.
The model potential is Eq.~(\ref{parameterization}) with $\kappa=0.5$.
The non-interacting problem has one band for $t_2/t_1 < 0.5$ and
two bands for $t_2/t_1 > 0.5$, cf.\ Fig.~\ref{energy dispersion},
and we focus on the latter region.
The limit $\gamma \to \infty$ corresponds to the Hubbard model with
on-site repulsion only, and the C2S2 phase is stable only over a
narrow window $t_2/t_1 \in [0.5 \dots 0.57]$.\cite{Louis01, Balents96}
The C2S2 region becomes progressively wider as we increase the
interaction range $1/\gamma$.}
   \label{extendedc2s2}
\end{figure}

The extended repulsion, Eq.~(\ref{parameterization}), is in Fourier space
\begin{eqnarray}
\label{extendedrepul}
V_Q = U
\left[1 - \kappa
      + \frac{\kappa \sinh(\gamma)}{\cosh(\gamma) - \cos(Q)} \right].
\end{eqnarray}
For given model parameters, we use
Eqs.~(\ref{rho11})-(\ref{wsigma12}) to set initial conditions.
We follow the RG flows and identify the phases as described above,
thus mapping out the ``weak coupling phase diagram''.
Here and in the rest of the paper, we take $\kappa = 0.5$.  This is
loosely motivated by the recent ab initio calculation\cite{Nakamura09}
for the \ET\ which gives the ratio of the nearest neighbor repulsion
$V_1 \equiv V(|x-x'|=1)$ to the on-site Hubbard $U$ as
$V_1/U \simeq 0.43$, while in our model $V_1 / U = \kappa e^{-\gamma}$.
The corresponding phase diagram showing stable C2S2 region is in
Fig.~\ref{extendedc2s2}.

We see that the C2S2 region becomes wider upon increasing the
interaction range $1/\gamma$.  We can understand this qualitatively
as follows.  For fixed band parameters, when $\gamma \to 0$ the
values of $V_Q$ for all non-zero $Q$ approach $U(1-\kappa)$, while
$V_{Q=0} \simeq 2 \kappa U / \gamma$ continues to increase.
The corresponding contribution to $\Lambda^\rho$ is
\begin{eqnarray}
\nonumber \delta\Lambda^\rho &=&
\frac{V_{Q=0}}{2\pi} \left[\frac{1}{v_1} + \frac{1}{v_2}
- \frac{4}{v_1 + v_2} \right] \\
&=& \frac{V_{Q=0}}{2\pi}\left[\frac{( v_1-v_2 )^2}{v_1 v_2 (v_1+v_2)}\right]~,
\end{eqnarray}
which is positive for any $v_1 \neq v_2$ and grows with increasing
$V_{Q=0}$.  Note also from Eqs.~(\ref{rho11})-(\ref{wsigma12}) that the
$V_{Q=0}$ enters only in the $\lambda_{ab}^\rho$ couplings.
Large bare value of $\Lambda^\rho$ makes the $w_{12}^{\rho/\sigma}$ flows
strongly irrelevant.  Their effect on the $\lambda^\sigma_{ab}$ flows is
rapidly decreasing and expires.  The $\lambda^\sigma_{ab}$ couplings start
repulsive and stay so and eventually flow to zero via marginal flows.
This argument is strictly true in the small $\gamma$ limit,
while for finite $\gamma$ the interplay of different flows is more
complex and requires numerical study as done in Fig.~\ref{extendedc2s2}.

\subsubsection{Weak coupling phase diagram for potential
Eq.~(\ref{parameterization}) truncated at the 4-th neighbor}
\label{subsubsec:truncatedc2s2}

\begin{figure}[t]
   \centering
   \includegraphics[width=\columnwidth]{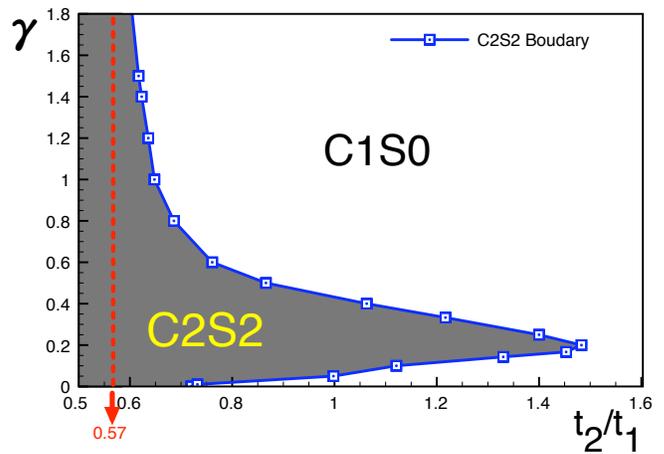}
   \caption{
Same as Fig.~\ref{extendedc2s2} but for the potential
Eq.~(\ref{parameterization}) truncated at the 4-th neighbor.}
   \label{truncatedc2s2}
\end{figure}

Here, we truncate the interaction at the 4-th neighbor,
so the Fourier transform is,
\begin{eqnarray}
\nonumber
V_Q = U \left[1 + 2 \kappa \sum_{n=1}^4 e^{-n\gamma} \cos(nQ) \right].
\end{eqnarray}
The phase diagram in the weak coupling RG approach is shown in
Fig.~\ref{truncatedc2s2}.
We see that unlike the case without the truncation, the C1S0 phase
opens again as $\gamma \to 0$.
Since we only include up to the 4-th neighbor interaction, $V_{Q=0}$
does not dominate over $V_{Q\ne 0}$ even in the $\gamma \to 0$
limit.  For $\kappa = 0.5$ and $\gamma = 0$, there is significant
structure in $V_Q$ including sign changes as a function of $Q$,
which can make bare spin couplings $\lambda_{aa}^\sigma \sim V_{2k_{Fa}}$
to be marginally relevant.
Nevertheless, for intermediate $\gamma$ there is still a wide window
of the C2S2 phase.

\section{Weak to intermediate coupling: phases out of C2S2 upon
increasing interaction}
\label{sec:Intermediate coupling}

\subsection{Harmonic description of the C2S2 phase}

Let us begin with a harmonic description of the C2S2 metal.
Technical steps and many details of the bosonization essentially
follow Ref.~\onlinecite{Sheng09} and references therein.
We write
\begin{equation}
c_{P a \alpha} = \eta_{a\alpha}
e^{ i (\varphi_{a\alpha} + P \theta_{a\alpha}) } ~,
\label{cbosonize}
\end{equation}
where $\varphi$ and $\theta$ are canonically conjugate boson fields
and $\eta$ are Klein factors.

We define ``charge" and ``spin" boson fields,
\begin{equation}
\theta_{a \rho /\sigma} = \frac{1}{\sqrt{2}}
(\theta_{a \up} \pm \theta_{a \dn}) ~,
\end{equation}
and ``even" and ``odd" flavor combinations,
\begin{equation}
\theta_{\mu \pm} = \frac{1}{\sqrt{2}}
(\theta_{1\mu} \pm \theta_{2\mu}) ~,
\end{equation}
with $\mu = \rho, \sigma$.  Similar definitions hold for the
$\varphi$ fields.

We can now bosonize all four-fermion interactions
Eqs.~(\ref{rlrhohamiltonian})-(\ref{chiralsigmahamiltonian}).
First consider the spin sector.  The $C^\sigma_{ab}$ terms give
velocity renormalizations, while the $\lambda^\sigma_{ab}$ terms
are written out in Sec.~IVA of Ref.~\onlinecite{Sheng09} and are
not repeated here.  We assume that the $\lambda^\sigma_{ab}$ are
marginally irrelevant in the C2S2 phase.
The fixed point Lagrangian has effectively decoupled boson fields
$\theta_{1\sigma}$ and $\theta_{2\sigma}$ with Luttinger parameters
$g_{1\sigma} = g_{2\sigma} = 1$, dictated by SU(2) spin rotation
invariance.

The Lagrangian in the charge sector is
\begin{eqnarray}
\mathcal{L}^\rho &=& \frac{1}{2\pi}
\left[\partial_x \bm{\Theta}^T \cdot {\bf A} \cdot \partial_x \bm{\Theta}
      + \partial_x \bm{\Phi}^T \cdot {\bf B} \cdot \partial_x \bm{\Phi}
\right] \nonumber \\
&& + \frac{i}{\pi} \partial_x \bm{\Theta}^T \cdot \partial_\tau \bm{\Phi}
~,
\label{Lrho}
\end{eqnarray}
where we defined $\bm{\Theta}^T = (\theta_{\rho+}, \theta_{\rho-})$ and
$\bm{\Phi}^T = (\varphi_{\rho+}, \varphi_{\rho-})$.
Matrix elements of ${\bf A}$ and ${\bf B}$ are:
\begin{widetext}
\begin{eqnarray}
A_{11} &=& \bar{v}
+ \frac{\lambda^\rho_{11} + \lambda^\rho_{22} + 2\lambda^\rho_{12}}{2\pi}
+ \frac{C^\rho_{11} + C^\rho_{22} + 2 C^\rho_{12}}{2\pi}
= \bar{v} + \frac{4 V_{Q=0}}{\pi}
- \frac{V_{2 k_{F1}}}{4\pi} - \frac{V_{2 k_{F2}}}{4\pi} - \frac{V_{\pi/2}}{2\pi}
- \frac{V_{k_{F1} - k_{F2}}}{2\pi} - \frac{U}{2\pi} ~,
\label{matrixelements:A11} \\
A_{22} &=& \bar{v}
+ \frac{\lambda^\rho_{11} + \lambda^\rho_{22} - 2\lambda^\rho_{12}}{2\pi}
+ \frac{C^\rho_{11} + C^\rho_{22} - 2 C^\rho_{12}}{2\pi}
= \bar{v}
- \frac{V_{2 k_{F1}}}{4\pi} - \frac{V_{2 k_{F2}}}{4\pi} + \frac{V_{\pi/2}}{2\pi}
+ \frac{V_{k_{F1} - k_{F2}}}{2\pi} - \frac{U}{2\pi} ~,
\label{matrixelements:A22} \\
A_{12} &=& A_{21} = v_r
+ \frac{\lambda^\rho_{11} - \lambda^\rho_{22}}{2\pi}
+ \frac{C^\rho_{11} - C^\rho_{22}}{2\pi}
= v_r - \frac{V_{2 k_{F1}}}{4\pi} + \frac{V_{2 k_{F2}}}{4\pi} ~,
\label{matrixelements:A12} \\
B_{11} &=& \bar{v}
- \frac{\lambda^\rho_{11} + \lambda^\rho_{22} + 2\lambda^\rho_{12}}{2\pi}
+ \frac{C^\rho_{11} + C^\rho_{22} + 2 C^\rho_{12}}{2\pi}
= \bar{v}
+ \frac{V_{2 k_{F1}}}{4\pi} + \frac{V_{2 k_{F2}}}{4\pi} + \frac{V_{\pi/2}}{2\pi}
- \frac{V_{k_{F1} - k_{F2}}}{2\pi} - \frac{U}{2\pi} ~,
\label{matrixelements:B11} \\
B_{22} &=& \bar{v}
- \frac{\lambda^\rho_{11} + \lambda^\rho_{22} - 2\lambda^\rho_{12}}{2\pi}
+ \frac{C^\rho_{11} + C^\rho_{22} - 2 C^\rho_{12}}{2\pi}
= \bar{v}
+ \frac{V_{2 k_{F1}}}{4\pi} + \frac{V_{2 k_{F2}}}{4\pi} - \frac{V_{\pi/2}}{2\pi}
+ \frac{V_{k_{F1} - k_{F2}}}{2\pi} - \frac{U}{2\pi} ~,
\label{matrixelements:B22} \\
B_{12} &=& B_{21} = v_r
- \frac{\lambda^\rho_{11} - \lambda^\rho_{22}}{2\pi}
+ \frac{C^\rho_{11} - C^\rho_{22}}{2\pi}
= v_r + \frac{V_{2 k_{F1}}}{4\pi} - \frac{V_{2 k_{F2}}}{4\pi} ~,
\label{matrixelements:B12}
\end{eqnarray}
\end{widetext}
where
\begin{eqnarray}
\bar{v} \equiv \frac{v_1 + v_2}{2}, \quad
v_r \equiv \frac{v_1 - v_2}{2} ~.
\end{eqnarray}
The couplings $\lambda^\rho_{ab}$ of the right-left mixing interactions
$\mathcal{H}^\rho_{RL}$ enter with opposite signs in ${\bf A}$ and
${\bf B}$ and directly affect Luttinger parameters, while the couplings
$C^\rho_{ab}$ of $\mathcal{H}^{\rho}_{\rm chiral}$ enter with the
same sign and give velocity renormalizations.

From the final expressions in terms of $V_Q$, we see that the $Q=0$
component enters only in $A_{11}$.  This can be understood by
considering the $Q=0$ part of the interaction,\cite{Kane97}
\begin{eqnarray}
\sum_{x,x'} V(x-x') n(x) n(x')
\;\to\; V_{Q=0} \int_x [\rho(x)]^2
\label{bosonize_V_Q0}
\end{eqnarray}
where $\rho(x) = 2\partial_x \theta_{\rho+}/\pi$ is the coarse-grained
electron density.

Note also that the $-U/(2\pi)$ in the diagonal matrix elements is
due to our more careful treatment of the on-site repulsion,
which we first write as $U n_\up(x) n_\dn(x)$ and then bosonize.

We obtain harmonic description of the C2S2 phase by combining the
spin and charge sectors.  The latter two-mode system $\mathcal{L}^\rho$
has nontrivial Luttinger parameters, which can be determined from the
matrices ${\bf A}$ and ${\bf B}$ (cf.~Appendix~\ref{app:scalingdimc2s2}).
The fixed-point matrix elements will differ somewhat from the bare
values above, but we ignore this in our crude analysis of the
intermediate coupling regime.

To complete the bosonization of the four-fermion interactions,
Eqs.~(\ref{rlrhohamiltonian})-(\ref{chiralsigmahamiltonian}),
the $w^{\rho/\sigma}_{12}$ terms give~\cite{Lin97, Sheng09}
\begin{eqnarray}
&&\hspace{-1cm}
W \equiv \left(w^\rho_{12} J_{R12} J_{L12}
               - w^\sigma_{12} \vec{J}_{R12} \cdot \vec{J}_{L12} \right)
+ \Hc \\
&&\hspace{-1cm}
= \cos(2\varphi_{\rho-}) \Bigg\{
4 w^\rho_{12} \left[\cos(2\varphi_{\sigma-})
                    - \hat{\Gamma} \cos(2\theta_{\sigma-}) \right]
\nonumber \\
&&\hspace{-1cm}
- w^\sigma_{12} \left[\cos(2\varphi_{\sigma-})
                      + \hat{\Gamma} \cos(2\theta_{\sigma-})
                      + 2 \hat{\Gamma} \cos(2\theta_{\sigma_+}) \right]
\Bigg\} ~,
\label{Wterm}
\end{eqnarray}
where $\hat{\Gamma} = \eta_{1\up} \eta_{1\dn} \eta_{2\up} \eta_{2\dn}$.
We see that $W$ couples the charge and spin sectors.
In the C2S2 theory described above, its scaling dimension is,
\begin{eqnarray}
\Delta[W] = \Delta[\cos(2\varphi_{\rho-})] + 1 ~,
\label{dimW}
\end{eqnarray}
where $\Delta[\cos(2\varphi_{\rho-})]$ is evaluated in the Lagrangian
$\mathcal{L}^\rho$, while the contribution $1$ comes from the
spin sector.  For the C2S2 theory to be consistent, the $W$ term must
be irrelevant, $\Delta[W] > 2$.  Once the $W$ renormalizes to zero,
the charge and spin sectors decouple.
We thus have precise parallel with the weak coupling analysis of the
C2S2 fixed point in Sec.~\ref{sec:Weak coupling}.

On the other hand, if $\Delta[W] < 2$, the $W$ term becomes relevant
and the C2S2 state is unstable.  In this case, $\varphi_{\rho-}$ will
get pinned and also the spin sector will become gapped.
Only the ``$\rho+$'' mode remains gapless and the system is some
C1S0 conducting phase.

\subsection{Mott insulator driven by Umklapp interaction.
Intermediate coupling procedure out of the C2S2}

The weak coupling analysis in Sec.~\ref{sec:Weak coupling} misses the
possibility of gapping out the overall charge mode $\theta_{\rho+}$
since there are no four-fermion Umklapp terms allowed in the
two-band system.  However, the half-filled electronic system does
become a Mott insulator for sufficiently strong repulsion.
In the theoretical description, this is achieved by an
eight-fermion Umklapp term\cite{Sheng09}
\begin{eqnarray}
H_8 &=& v_8
(c_{R1\up}^\dagger c_{R1\dn}^\dagger c_{R2\up}^\dagger c_{R2\dn}^\dagger
 c_{L1\up} c_{L1\dn} c_{L2\up} c_{L2\dn} + \Hc ) \nonumber \\
&=& 2 v_8 \cos(4 \theta_{\rho+}) ~.
\label{H8}
\end{eqnarray}
At weak coupling, this term has scaling dimension $\Delta[H_8] = 4$
and is strongly irrelevant.
However, from Eq.~(\ref{bosonize_V_Q0}) we see that overall repulsive
interaction stiffens the $\theta_{\rho+}$ mode and lowers the
scaling dimension of $H_8$.  For sufficiently strong repulsion,
$\Delta[H_8]$ drops below $2$ and the Umklapp becomes relevant;
$\theta_{\rho+}$ gets pinned and we obtain a Mott insulator.

Our intermediate coupling procedure is as follows.
Using the harmonic theory of the C2S2 phase, we calculate the
scaling dimensions $\Delta[W]$, Eq.~(\ref{dimW}), and
$\Delta[H_8] = \Delta[\cos(4\theta_{\rho_+})]$
from the Lagrangian $\mathcal{L}^\rho$, Eq.~(\ref{Lrho}).
Details are described in Appendix~\ref{app:scalingdimc2s2}
and calculations are done numerically in the end.

If both $\Delta[W]$ and $\Delta[H_8]$ are larger than $2$, the
C2S2 metal is stable.  As interactions increase, eventually
either $W$ or $H_8$ becomes relevant.
In general, there are two cases:

1) If $H_8$ becomes relevant first, we pin $\theta_{\rho+}$ and enter
``C1S2'' Mott insulator.  To be more precise, we can further qualify
the label as ``C1[$\rho-$]S2''; the remaining ``charge'' mode
``$\rho-$'' represents local current loop fluctuations and
does not conduct.
This is the spin liquid phase called Spin Bose-metal in
Ref.~\onlinecite{Sheng09} and described in detail there.
Exploring conditions for finding such phase in electronic models
is our main goal here.

2) On the other hand, if the $W$ term becomes relevant first,
we enter C1S0 conducting state with a spin gap
(more precisely, ``C1[$\rho+$]S0'').

Some reservations are in order.
First, we use bare values of the couplings in the ${\bf A}$ and ${\bf B}$
matrices, which is not accurate since the couplings experience
initial flows, cf.\ Sec.~\ref{subsec:RGeqs}.
Second, we consider only instabilities driven by changes in the
harmonic ${\cal L}^\rho$ theory as they translate to scaling dimensions
of the $H_8$ and $W$ terms, i.e., we effectively treat the latter as
small.
We also assume that the spin sector is near the fixed point with all
$\lambda^\sigma_{ab}$ marginally irrelevant and small.
We will address these reservations after presenting results of the
above procedure.
Keeping these remarks in mind, we now describe how we analyze phases
out of the C1[$\rho-$]S2 and C1[$\rho+$]S0 in the same procedure.

\subsubsection{Instability out of C1[$\rho-$]S2 driven by spin-charge
coupling $W$}
In the present analysis focusing on the ``$\rho+$'' and ``$\rho-$''
fields, we can also crudely estimate the extent of the C1S2 or C1S0
phases once either happens out of the C2S2.

Suppose the Umklapp $H_8$ is relevant first and we are in the C1S2 phase.
We still need to remember the $W$ term since it can become relevant
if we continue increasing the interaction strength.
To estimate the scaling dimension of the $W$ term, we assume now that
the $\theta_{\rho+}$ field is massive and integrate out $\theta_{\rho+}$
and $\varphi_{\rho+}$.
Mathematically this amounts to sending $A_{11} \to \infty$, and we obtain
\begin{eqnarray}
\Delta\left[ W; ~\theta_{\rho+}~\text{is pinned} \right] =
\left[ \frac{A_{22} B_{11}}{B_{11} B_{22} - B_{12}^2}
\right]^{\frac{1}{2}} + 1 ~.
\label{dimW_in_C1S2}
\end{eqnarray}
This assumption is approximate but reasonable, since once the parameters
are such that the system is in the C1S2 phase, the relevant $H_8$ will
grow and quickly stiffen the $A_{11}$ in positive feedback loop.

The C1S2 phase is stable if $\Delta[W] > 2$, and this analysis
is similar to the stability analysis of the SBM in
Ref.~\onlinecite{Sheng09}.
If $\Delta[W]$ drops below 2, the $W$ term becomes relevant and the
$\varphi_{\rho-}$ field will be pinned, together with gapping out the
spin sector, cf.\ Eq.~(\ref{Wterm}).  The final result is some
``C0S0'' phase, whose precise character depends on the details of the
couplings $w_{12}^{\rho/\sigma}$.
This is studied in Sec.~IVB of Ref.~\onlinecite{Sheng09}.
For the present repulsive electron model, we have
$w_{12}^\rho, w_{12}^\sigma > 0$, so the resulting C0S0 is likely a
period-2 Valence Bond Solid (VBS).\cite{Sheng09}
This connects to dimerized phase in the $J_1 - J_2$ spin chain
appropriate in the strong interaction limit of the electron system.

\subsubsection{Instability out of C1[$\rho+$]S0 driven by Umklapp
$H_8$}

Suppose now the $W$ interaction becomes relevant first.
From Eq.~(\ref{Wterm}), it is natural that $\varphi_{\rho-}$ is pinned,
the spin sector gets gapped, and we are in C1S0 phase.
Here we postulate mass for $\varphi_{\rho-}$
(essentially sending $B_{22} \to \infty$) and calculate the effective
scaling dimension of the Umklapp term,
\begin{eqnarray}
\Delta\left[ H_8; ~\varphi_{\rho-}~\text{is pinned} \right] =
4 \left[ \frac{B_{11} A_{22}}{A_{11} A_{22} - A_{12}^2}
  \right]^{\frac{1}{2}} ~.
\label{dimUmklapp_in_C1S0}
\end{eqnarray}
If $\Delta[H_8] > 2$, the C1S0 is stable.  Once $\Delta[H_8]$ drops
below $2$, the overall charge mode $\theta_{\rho+}$ is pinned and we
obtain fully gapped Mott insulator C0S0, which is likely the same
period-2 VBS discussed earlier.

\subsection{Numerical results}
We consider the same models with extended density-density interactions
as in the weak coupling analysis in Sec.~\ref{subsec:RG:numphased},
parking ourselves initially in the C2S2 phase in Fig.~\ref{extendedc2s2}
and Fig.~\ref{truncatedc2s2}.
From the preceding discussion, we can obtain two phases out of the C2S2
upon increasing interaction strength -- either C1[$\rho+$]S0 or
C1[$\rho-$]S2.
To visualize the results, we imagine adding the overall interaction
strength $V$ as the $z$-axis to Fig.~\ref{extendedc2s2} and
Fig.~\ref{truncatedc2s2}.  We then project down which phase happens first
for each such vertical line out of C2S2.  Calculations are done
numerically and the results are shown in Fig.~\ref{projectionphase}
and Fig.~\ref{projectionphasetrun}.
In Fig.~\ref{gamma=0.4} we take a cut through Fig.~\ref{projectionphase}
at $\gamma=0.4$ and show details of the phase diagram in the
$t_2/t_1 - V$ plane.

\subsubsection{Intermediate coupling phase diagram for model with
potential Eq.~(\ref{parameterization})}

\begin{figure}[t]
  \centering
  \includegraphics[width=\columnwidth]{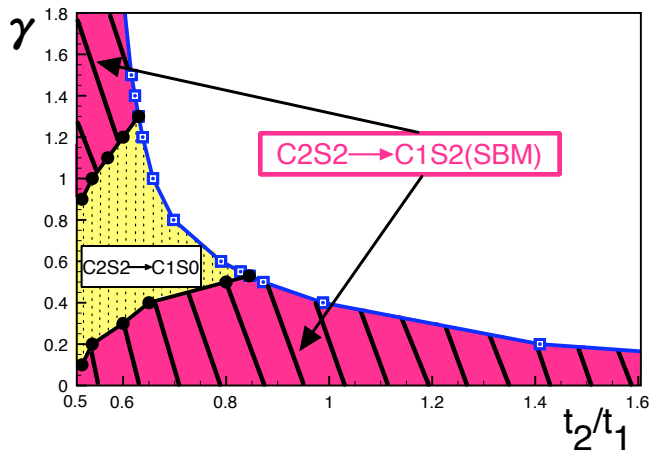}
  \caption{
Projection of phases obtained out of the C2S2 of Fig.~\ref{extendedc2s2}
as we increase overall repulsion strength $V$,
which we imagine to be the $z$-axis perpendicular to the page
(Fig.~\ref{gamma=0.4} gives one cut at $\gamma=0.4$ with such $V$ axis
shown explicitly).
The results are obtained in the intermediate coupling procedure as
explained in the text.  White region is C1S0 at weak coupling,
cf.~Fig.~\ref{extendedc2s2}, and is not considered here.
}
  \label{projectionphase}
\end{figure}

Fig.~\ref{projectionphase} shows results for the model potential
Eq.~(\ref{parameterization}).  We can see that in two regimes
$\gamma \geq 1.2$ and $\gamma \leq 0.4$ we exit from the C2S2
into the C1S2.  The two limits can be understood analytically.

In the large $\gamma$ case, we can replace all $V_Q$ by simply $U$.
The matrices ${\bf A}$ and ${\bf B}$ defined in
Eq.~(\ref{matrixelements:A11})-(\ref{matrixelements:B12}) become
\begin{eqnarray}
{\bf A} = \begin{pmatrix}
\bar{v} + \frac{2 U}{\pi} & v_r \\
v_r & \bar{v}
\end{pmatrix} ~, \quad
{\bf B} = \begin{pmatrix}
\bar{v} & v_r \\
v_r & \bar{v}
\end{pmatrix} ~.
\end{eqnarray}
We see that $U$ only contributes to $A_{11}$.  This monotonically
``stiffens'' the $\theta_{\rho+}$ (lowering $\Delta[H_8]$)
but ``softens'' the $\varphi_{\rho-}$ (increasing $\Delta[W]$).
Therefore we only expect the C1S2 phase out of the C2S2 as found in the
numerical calculations.

On the other hand, for small $\gamma$ we can see from
Eq.~(\ref{extendedrepul}) that $V_{Q=0}$ will dominate over $V_{Q\neq 0}$.
Keeping only $V_{Q=0}$, the matrices ${\bf A}$ and ${\bf B}$ become
\begin{eqnarray}
{\bf A} \simeq \begin{pmatrix}
\bar{v} + \frac{4 V_{Q=0}}{\pi} & v_r \\
v_r & \bar{v}
\end{pmatrix} ~, \quad
{\bf B} \simeq \begin{pmatrix}
\bar{v} & v_r \\
v_r & \bar{v}
\end{pmatrix} ~.
\end{eqnarray}
Thus the small $\gamma$ case has similar mathematical structure to the
large $\gamma$ case.  The physical difference is that here the
transition to the C1S2 is driven by the $V_{Q=0}$ instead of the
on-site Hubbard $U$.
Note also that since $V_{Q=0} \simeq 2\kappa U/\gamma$ for
$\gamma \ll 1$, the transition requires only small values of $U$,
which is why we can ignore all $V_{Q \neq 0}$ compared to the band
velocities.

\begin{figure}[t]
\centering
  \includegraphics[width=\columnwidth]{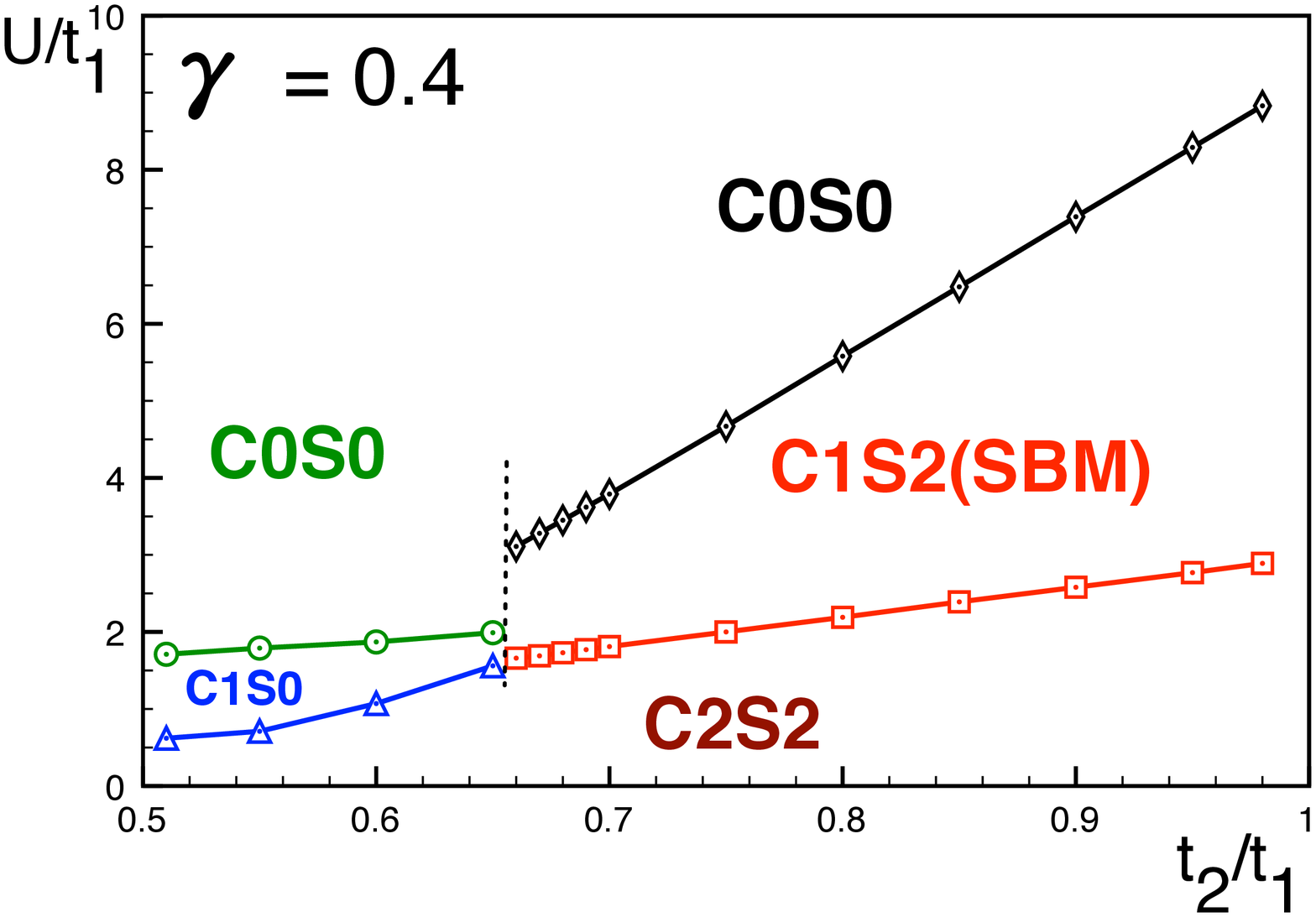}
  \caption{
(Color online)
Intermediate coupling analysis of the model with potential
Eq.~(\ref{parameterization}) for $\kappa = 0.5$ and $\gamma = 0.4$.
Here the horizontal range is equal to the extent of the C2S2 phase in
the weak coupling analysis from Fig.~\ref{extendedc2s2}.
We start in the C2S2 at small $U$.
The boundary where the charge-spin coupling term $W$ becomes relevant
first is indicated with blue triangles and the system goes into the C1S0;
the next stage where the C1S0 in turn becomes unstable and the
system goes into the C0S0 is marked with green circles.
The boundary where the Umklapp term $H_8$ becomes relevant first is
indicated with red squares and the system goes into the C1S2,
which is the SBM phase of Ref.~\onlinecite{Sheng09};
upon further increase of the interaction strength the C1S2 eventually
becomes unstable and goes to the C0S0 at locations marked with black
diamonds.
Note that the discontinuity shown with dotted vertical line is
not meaningful and is due to our crude analysis performed separately
out of the C1S0 and C1S2; in either case, the final C0S0 is likely
the same phase.   Also note that the C1 mode content is distinct
in the C1[$\rho+$]S0 (conducting) and C1[$\rho-$]S2 (insulating)
cases and any transition between them is first order.
The C2S2 to C1S2 transition is Kosterlitz-Thouless-like.
}
\label{gamma=0.4}
\end{figure}

Now we consider a cut at $\gamma = 0.4$ to see more details in the
$t_2/t_1 - V$ plane.  The results are shown in Fig.~\ref{gamma=0.4}.
Compared with the two limits $\gamma \gg 1$ and $\gamma \ll 1$ above,
all possibilities that we discussed out of the C2S2 are realized here.
The C1S0 phase appears for $t_2 / t_1 < 0.65$ for some quantitative
reasons.
Various $V_Q$ are all of the same order, unlike the $\gamma \ll 1$ case.
At the same time, they have some non-trivial $Q$-dependence, unlike the
$\gamma \gg 1$ case, which is somehow enough to make the $W$ term become
relevant and preempt the Umklapp term.
Note that for small interactions the scaling dimension of the $W$ term
can be obtained from the weak coupling RG equations for the
$w_{12}^{\rho/\sigma}$ in Sec.~\ref{subsec:RGeqs} by setting all
$\lambda^\sigma_{ab} = 0$ (since we ignore the spin sector
in the present procedure).  Thus, $\Delta[W] = 2 + \Lambda^\rho$,
where $\Lambda^\rho$ is defined in Eq.~(\ref{c2s2instability}).
Since $\Lambda^\rho$ can only decrease under the weak coupling RG
and the shaded C2S2 region in Fig.~\ref{extendedc2s2} was found to be
stable, we expect $\Delta[W]$ here to increase with $V$ for small $V$,
in agreement with numerical calculations.
However, we find that $\Delta[W]$ eventually starts to decrease with
increasing $V$ and can become relevant before the Umklapp.  This is a
quantitative matter and comes from putting together all interactions
$\mathcal{H}^\rho_{RL}$ and $\mathcal{H}^\rho_{\rm chiral}$,
Eq.~(\ref{rlrhohamiltonian})-(\ref{chiralrhohamiltonian}),
in the intermediate coupling procedure.  Such numerical calculations
give us that the C2S2 can exit into the C1S0 phase.
For larger $t_2 / t_1 > 0.65$ in Fig.~\ref{gamma=0.4}, we obtain
the sought for C1S2 spin liquid phase.

This concludes the presentation of formal results within the
particular procedure for intermediate scale analysis.
Let us now think how to combine the weak and intermediate coupling
approaches more realistically and see where our results are
more robust.

\begin{figure}[t]
  \centering
  \includegraphics[width=\columnwidth]{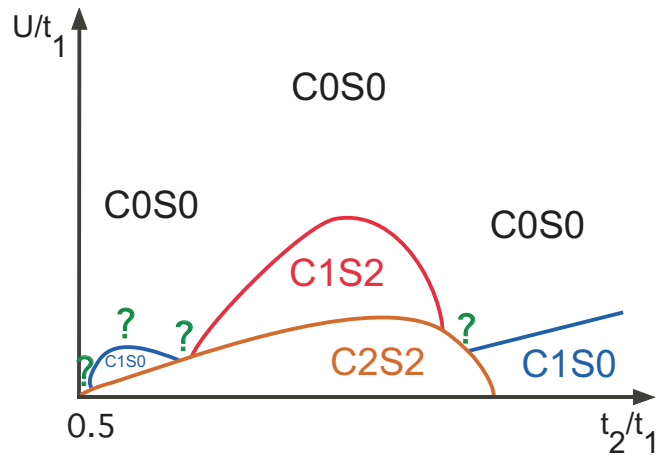}
  \caption{
Schematic merging of the weak and intermediate coupling results
in the model regimes like in Fig.~\ref{gamma=0.4} in the whole range
with $t_2/t_1 > 0.5$.
In weak coupling, the C2S2 phase is unstable beyond the shaded region
in Fig.~\ref{extendedc2s2}.
However, due to the crudeness of our intermediate coupling procedure,
Fig.~\ref{gamma=0.4} shows monotonic growth of the C2S2 phase with
$t_2/t_1$ past this instability.
This discrepancy arises because our intermediate coupling procedure
completely ignores the spin sector.
More realistically, we expect the C2S2 phase to peak somewhere in the
middle of the range shown in Fig.~\ref{gamma=0.4} and be bounded by
the C1S0 for larger $t_2/t_1$.
Similar considerations apply to the C1S2 phase, which is bounded
by the C0S0.
}
  \label{schematic phase}
\end{figure}

First of all, in the weak coupling analysis the C2S2 phase is
unstable beyond the shaded regions in Figs.~\ref{extendedc2s2} and
\ref{truncatedc2s2}.  However, this is lost in the specific
intermediate coupling procedure, which, when applied for small coupling,
would give C2S2 essentially everywhere.
For example, in Fig.~\ref{gamma=0.4} we see monotonic growth of the
C2S2 phase with $t_2/t_1$ past the point where the weak coupling analysis predicts instability.
The reason for this discrepancy is the complete neglect of the
spin sector in the formal intermediate scale procedure.
Indeed, in the weak coupling analysis, the instabilities manifest
dramatically once one of the $\lambda^\sigma_{aa}$ becomes negative,
causing runaway flows.  This can happen even when the bare
$\lambda^\sigma_{aa}$ are repulsive because they are renormalized
downwards and can be driven negative by the $w_{12}^{\rho/\sigma}$
contributions in Eqs.~(\ref{rglambdasigma11})-(\ref{rglambdasigma22}),
where we assume $w_{12}^\rho w_{12}^\sigma > 0$.
Also, the $\lambda^\sigma$ couplings feed back into the flow of
$w_{12}^{\rho/\sigma}$, so the RG flow behavior is even more complex.
So far we have dealt with this inadequacy of the intermediate scale
procedure by simply cutting it at the C2S2 boundaries determined from
the weak coupling analysis.  More realistically, we expect the
extent of the C2S2 phase to peak somewhere in the middle of the range
shown in Fig.~\ref{gamma=0.4} and decrease towards the right boundary.
Similar considerations apply to the C1S2 phase, which is likely confined
within the same $t_2/t_1$ range as the C2S2.
Therefore, the $t_2/t_1 - U/t_1$ phase diagram should be more like
Fig.~\ref{schematic phase}.

We can also discuss our earlier reservation about using bare values of
the couplings instead of some renormalized values.
Thinking about some RG treatment, we expect that crude patterns of
how various couplings affect each other are likely similar at
intermediate and weak couplings.  Now if we formally take the flow
equations from Sec.~\ref{subsec:RGeqs}, the outcome does not depend on
the initial interaction scale, so we would conclude the C2S2 phase
throughout the shaded region in Figs.~\ref{extendedc2s2}.
The weak coupling flow equations miss velocity renormalizations
due to chiral interactions, but these are not expected to flow strongly
and are treated reasonably in the intermediate coupling analysis.
The fact that the couplings are now finite and comparable with bare
band energies is also treated reasonably at intermediate coupling
due to the power of bosonization, so the outlined forging of weak and
intermediate scales seems appropriate.
Finally, the Umklapp term that is missing in the weak coupling approach
will feed into stiffening of $\theta_{\rho+}$ only, which is good for
the first instability out of the C2S2 to be into the C1S2 spin liquid.

We think that our conclusions are more robust for small $\gamma$ where
the extent of the C2S2 phase is larger and also the longer-ranged
potential is feeding precisely into stiffening the overall charge field
$\theta_{\rho+}$, which is good for going to the C1S2 phase.
On the other hand, results at medium to large $\gamma$ are likely
less reliable, with different scenarios depending on quantitative issues.

\subsubsection{Intermediate coupling phase diagram for model with
potential Eq.~(\ref{parameterization}) truncated at the 4-th neighbor}

\begin{figure}[t]
  \centering
  \includegraphics[width=\columnwidth]{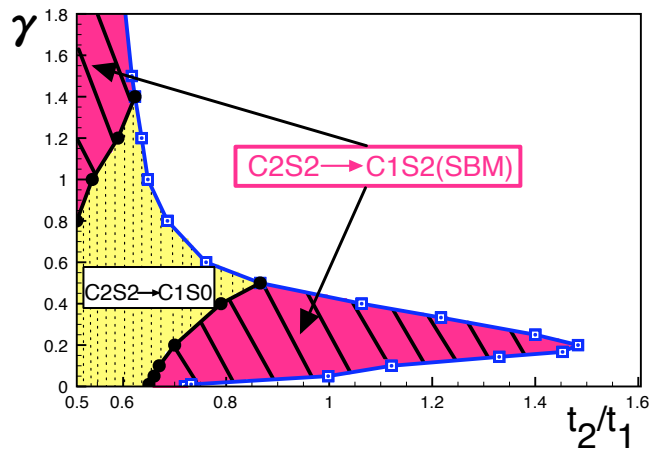}
  \caption{
Same as Fig.~\ref{projectionphase} but for the model with interactions
Eq.~(\ref{parameterization}) truncated at the 4-th neighbor
and starting out of the C2S2 of Fig.~\ref{truncatedc2s2}.
}
  \label{projectionphasetrun}
\end{figure}

Figure~\ref{projectionphasetrun} shows results of the intermediate
coupling analysis for the model with interactions truncated
at the 4-th neighbor, cf.~Sec.~\ref{subsubsec:truncatedc2s2}.
We have a rather similar story to Fig.~\ref{projectionphase},
except that the initial C2S2 region is bounded.
Large part of the C2S2 phase exits into the C1S2 spin liquid
upon increasing interactions, and our results are probably more
robust near $\gamma \sim 0.2 - 0.3$ where the C2S2 has the largest
extent along the $t_2/t_1$ axis.

\section{Summary and discussion}\label{conclusion}
To summarize, in this paper we consider {\it electronic} models for
realizing Spin Bose-metal (spin liquid) phase on the 2-leg triangular
strip found in Ref.~\onlinecite{Sheng09} in spin-1/2 model with ring
exchanges.  We identify the SBM with the C1S2 Mott insulator of electrons.

In Sec.~\ref{sec:Weak coupling}, we start with a two-band electron
system, which is C2S2.  Instead of considering only the on-site
Hubbard-type repulsion,\cite{Louis01, Daul, Fabrizio97, Capello, Japaridze, Valenti09, Gros09}
we study generally longer ranged density-density repulsion.
This is motivated in part by the expectation that real Coulomb
interaction is not screened in Mott insulator materials, so further
neighbor repulsion can be significant, as brought up by recent
ab initio work\cite{Nakamura09} for the spin liquid material \ET.
Using weak coupling RG analysis for the zigzag chain problem,
\cite{Balents96, Fabrizio96, Lin97, Louis01}
we find that such extended interactions open much wider window of the
C2S2 metal compared with the Hubbard model.
The main results are shown in Fig.~\ref{extendedc2s2} and
Fig.~\ref{truncatedc2s2}.  In the first figure, we have essentially
an independent control over the $Q=0$ part of the potential by
allowing it to extend to far neighbors, and we identify the dominance
of $V_{Q=0}$ as the main stabilizing force for the metal.
In the second figure, we truncate interactions at the 4-th neighbor
to check the robustness of our conclusions, in view that such models
may be easier to explore using numerical DMRG.
Our detailed quasi-1D considerations agree with the intuition
that in real metals electronic pairing instabilities are suppressed
by the long-ranged piece of the Coulomb interaction.
Such widening of the C2S2 region by extending the model interaction range
is warranted if we want to bring the electronic ladder system
closer to realistic situations in the 2D candidate spin liquid
materials.

In Sec.~\ref{sec:Intermediate coupling}, we begin with stable C2S2
metal at weak coupling and use bosonization to extend the analysis to
intermediate coupling by gradually increasing the overall repulsion
strength.  Within effective bosonic theory, we identify potential
instabilities of the C2S2 phase to spin-charge interaction $W$
[Eq.~(\ref{Wterm})] and Umklapp interaction $H_8$ [Eq.~(\ref{H8})].
The $W$ can drive the system into C1[$\rho+$]S0 phase with spin gap
but still conducting along the chain, while the Umklapp $H_8$ can
produce C1[$\rho-$]S2 Mott insulator with three gapless modes,
which is the desired SBM phase.
We calculate the scaling dimensions of the $W$ and $H_8$ terms
in the harmonic theory of the C2S2 metal using bare couplings
in the charge sector and assuming stability in the spin sector
-- this constitutes our naive intermediate coupling procedure.
The calculation of scaling dimensions is described in
Appendix~\ref{app:scalingdimc2s2} and is done numerically in the end.

We consider two cases depending on which of the terms $W$ or $H_8$
becomes relevant first and apply similar intermediate coupling
approach inside the resulting phase.   Assuming strong field pinning
by the already relevant term, we calculate the scaling dimension of the
remaining term and estimate when it eventually drives the system into
fully gapped C0S0 paramagnet (which is likely connected to the
dimerized phase of the $J_1 - J_2$ Heisenberg model at strong coupling).
With the help of such admittedly crude analysis, we can map out the
phase diagram in weak to intermediate coupling regime as illustrated
in schematic Fig.~\ref{schematic phase} (based on more naive
Fig.~\ref{gamma=0.4}).
Figures~\ref{projectionphase} and \ref{projectionphasetrun}
summarize our results and show where the C2S2 metal goes to the
C1S2 (SBM spin liquid) upon increasing overall repulsion strength.
We conclude that the C1S2 phase is quite natural out of the wider C2S2
metallic region, in particular when driven by extended repulsive
interactions.
It would be very interesting to confront our theoretical predictions
with numerical DMRG studies of such electronic models with extended
repulsion.

So far, we have approached the intermediate coupling Mott insulator
from the weak coupling metallic side.  One could try to attack the
same problem starting from the strong coupling limit deep in the
Mott insulator where Heisenberg spin-1/2 model is appropriate.
As one nears the metallic phase, it becomes important to include
multiple spin exchanges in the effective spin Hamiltonian to better
capture charge fluctuations in the underlying electron system.
\cite{McDonald, ringxch}
This is the motivation behind Ref.~\onlinecite{Sheng09} studying
$J_1 - J_2$ chain with additional four-spin ring exchanges.
The concept study Ref.~\onlinecite{Sheng09} allowed arbitrary variation
of the ring coupling compared with the Heisenberg couplings.
However, coming from an electronic model these do not vary independently
and more exchange terms are also generated.  It would be interesting to
pursue such approach systematically studying effective spin models with
multi-spin exchanges for realistic electronic models to see if they
harbor the SBM phase.  We do not make such attempts here, but only
give few simple observations on how the derivation of the spin model is
modified in the presence of extended repulsion.

First of all, for the two-spin exchanges, the familiar Hubbard model
expression $J_{rr'} = 4 t_{rr'}^2/U$ is modified to
$J_{rr'} = 4 t_{rr'}^2/(V_0 - V_{r-r'})$.  The energy denominator
is not simply the on-site $U = V_0$ but also includes interaction
potential between the two sites $r$ and $r'$.
For example, Ref.~\onlinecite{Nakamura09} estimates
$V_1 / V_0 \approx 0.43$ for the \ET\ spin liquid material,
and this would significantly affect values of the exchange constants.
Energy denominators for all virtual processes are similarly affected
and take a form of a charging energy for the deviations from the
background.  Multi-spin exchange amplitudes are given by a product of
electron tunneling amplitudes for a given virtual path divided by
a product of such charging energies in intermediate states along the
path.  Thus, the multi-spin exchanges may in fact be relatively more
important in systems with extended interactions.

As an extreme example, imagine a very slow decrease of $V(r-r')$
up to some distance $R$ (and perhaps a faster drop thereafter).
Then all exchange loops up to such radius $R$ will have large amplitudes.
The multi-spin exchanges encode the underlying kinetic energy of
electrons, and our intuition is that this would like to retain some
itinerancy in the spin degrees of freedom even when the charges
are localized.  From such strong to intermediate coupling perspective,
it appears that extended interactions would tend to stabilize the SBM
spin liquid near the insulator-metal transition, similar to our
conclusion from the weak to intermediate coupling study in the
quasi-1D models in this paper.
It would be interesting to pursue such considerations more carefully
and in realistic electronic models.
We hope that our work will further stimulate numerical studies of such
models on ladders and in two dimensions.

\acknowledgments

We would like to thank M.~P.~A.~Fisher, I. Gonzalez, R. Melko, and
D.~N.~Sheng for useful discussions and M.~P.~A.~Fisher for
stimulating this work and critical reading of the manuscript.
This research is supported by the A.~P.~Sloan Foundation
and the National Science Foundation through grant DMR-0907145.

\appendix

\section{Derivation of $\Delta[\cos{(4 \theta_{\rho+})}]$ and
$\Delta[\cos{(2 \varphi_{\rho-})}]$ in C2S2 phase}
\label{app:scalingdimc2s2}

Equation~(\ref{Lrho}) gives quadratic Lagrangian for the charge sector.
First, we redefine the fields which still satisfy the same
commutation relations,
\begin{eqnarray}
\bm{\Theta} = {\bf S} \cdot \bm{\Theta}_1 ~, \quad\quad
\bm{\Phi} = {\bf S} \cdot \bm{\Phi}_1 ~.
\end{eqnarray}
Here {\bf S} is an orthogonal $2\times2$ matrix diagonalizing the
matrix {\bf A},
\begin{eqnarray}
{\bf S}^T \cdot {\bf A} \cdot {\bf S} =
\begin{pmatrix}
A_1 & 0\\
0 & A_2
\end{pmatrix}
\equiv {\bf A}_D ~.
\end{eqnarray}
The Lagrangian becomes,
\begin{eqnarray}
\nonumber
\mathcal{L}^\rho &=& \frac{1}{2\pi}
\left[\partial_x \bm{\Theta}_1^T \cdot {\bf A}_D \cdot \partial_x \bm{\Theta}_1
      + \partial_x \bm{\Phi}_1^T \cdot {\bf S}^T \cdot {\bf B} \cdot {\bf S} \cdot \partial_x \bm{\Phi}_1
\right] \\
&& + \frac{i}{\pi} \partial_x \bm{\Theta}_1^T \cdot \partial_\tau \bm{\Phi}_1 ~.
\end{eqnarray}
Define another set of conjugate fields,
\begin{eqnarray}
\bm{\Theta}_1 = \frac{1}{\sqrt{{\bf A}_D}} \cdot \bm{\Theta}_2 ~,
\quad\quad
\bm{\Phi}_1 = \sqrt{{\bf A}_D} \cdot \bm{\Phi}_2 ~.
\end{eqnarray}
We obtain,
\begin{eqnarray}
\nonumber
\mathcal{L}^\rho &=& \frac{1}{2\pi}
\left[\partial_x \bm{\Theta}_2^T \cdot \partial_x \bm{\Theta}_2
      + \partial_x \bm{\Phi}_2^T \cdot {\bf B}' \cdot \partial_x \bm{\Phi}_2
\right] \\
&& + \frac{i}{\pi} \partial_x \bm{\Theta}_2^T \cdot \partial_\tau \bm{\Phi}_2 ~,
\end{eqnarray}
where
\begin{equation}
{\bf B}' \equiv \sqrt{{\bf A}_D} \cdot {\bf S}^T \cdot {\bf B} \cdot {\bf S} \cdot \sqrt{{\bf A}_D} ~.
\end{equation}

We use the same trick to diagonalize matrix {\bf B}$^\prime$:
\begin{eqnarray}
\bm{\Theta}_2 = {\bf R} \cdot \bm{\Theta}_3 ~, \quad\quad
\bm{\Phi}_2 = {\bf R} \cdot \bm{\Phi}_3 ~,
\end{eqnarray}
where {\bf R} is an orthogonal matrix which satisfies,
\begin{eqnarray}
{\bf R}^T \cdot {\bf B}' \cdot {\bf R} =
\begin{pmatrix}
B'_1 & 0\\
0 & B'_2
\end{pmatrix}
\equiv {\bf B}'_D ~.
\end{eqnarray}
The Lagrangian becomes,
\begin{eqnarray}
\label{scaling dimension}
\nonumber
\mathcal{L}^\rho &=& \frac{1}{2\pi}
\left[\partial_x \bm{\Theta}_3^T \cdot \partial_x \bm{\Theta}_3
      + \partial_x \bm{\Phi}_3^T \cdot {\bf B}'_D \cdot \partial_x \bm{\Phi}_3
\right]\\
&& + \frac{i}{\pi} \partial_x \bm{\Theta}_3^T \cdot \partial_\tau \bm{\Phi}_3 ~.
\end{eqnarray}
Now we can calculate the scaling dimension of
$\cos{(4 \theta_{\rho+})}$ and $\cos{(2 \varphi_{\rho-})}$ from
Eq.~(\ref{scaling dimension}) through relations,
\begin{eqnarray}
\bm{\Theta} &=& {\bf S} \cdot \frac{1}{\sqrt{{\bf A}_D}} \cdot {\bf R} \cdot \bm{\Theta}_3 ~, \\
\bm{\Phi} &=& {\bf S} \cdot \sqrt{{\bf A}_D} \cdot {\bf R} \cdot \bm{\Phi}_3 ~,
\end{eqnarray}
and scaling dimensions of the final fields,
\begin{eqnarray}
\Delta[e^{i \bm{\Theta}_3}] = \frac{\sqrt{{\bf B}'_D}}{4}~, \quad\quad
\Delta[e^{i \bm{\Phi}_3}] = \frac{1}{4\sqrt{{\bf B}'_D}} ~,
\end{eqnarray}
where the right hand sides mean corresponding diagonal matrix elements.
Therefore, we find general form for the dimensions we are interested in,
\begin{eqnarray}
\nonumber
\Delta[\cos{(4 \theta_{\rho+})}] &=&
4 \sqrt{B'_1} \left(\frac{S_{11} R_{11}}{\sqrt{A_1}}
                    + \frac{S_{12} R_{21}}{\sqrt{A_2}} \right)^2 \\
&+& 4 \sqrt{B'_2} \left(\frac{S_{11} R_{12}}{\sqrt{A_1}}
                         + \frac{S_{12} R_{22}}{\sqrt{A_2}} \right)^2 ~,
\label{umklappc2s2} \\
\nonumber
\Delta[\cos{(2\varphi_{\rho-})}] &=&
\frac{\left(\sqrt{A_1} S_{21} R_{11} + \sqrt{A_2} S_{22} R_{21}\right)^2}
     {\sqrt{B'_1}} \\
&+&
\frac{\left(\sqrt{A_1} S_{21} R_{12} + \sqrt{A_2} S_{22} R_{22}\right)^2}
     {\sqrt{B'_2}} ~, \label{wtermc2s2}
\end{eqnarray}
where $S_{ab}$ and $R_{ab}$ are matrix elements of {\bf S} and {\bf R}.


\bibliography{biblio4zigzag}

\end{document}